\documentclass[
12pt,
showpacs,
aps,
prd,
nofootinbib,
floatfix,
amsmath,
amssymb]{revtex4-1}

\usepackage{graphicx}
\usepackage{dcolumn}
\usepackage{bm}
\usepackage{float}
\usepackage{color}
\usepackage{hyperref}
\usepackage{mathrsfs}
\usepackage{rotating}
\usepackage{epstopdf}

\newcommand{\be}{\begin{eqnarray}}
\newcommand{\ee}{\end{eqnarray}}
\newcommand{\nn}{\nonumber}
\newcommand{\yd}{{{Y^d_1}}}
\newcommand{\yydd}{{{Y^d_2}}}

\newcommand{\yu}{{{Y^u_1}}}
\newcommand{\yyuu}{{{Y^u_2}}}

\newcommand{\hggs}{\Phi_1}
\newcommand{\hggss}{\Phi_2}
\newcommand{\hgt}{\tilde{\Phi}_1}
\newcommand{\hgtt}{\tilde{\Phi}_2}
\def \GeV{\mathop{\mbox{\normalfont GeV}}\nolimits}
\def \Br{\mathop{\mbox{\normalfont BR}}\nolimits}

\def \Mm{\mathcal{M}}
\def \Aa{\mathcal{A}}

\def \Gg{\mathcal{G}}
\def \dL{\mathcal{L}}

\begin{document}

\title{
One-loop decays ${A^0} \to ZZ, Z\gamma, \gamma\gamma$
within the 2HDM and its search at the LHC}

\author{J. L. Diaz-Cruz}
\email{jldiaz@fcfm.buap.mx}
\affiliation{Facultad de Ciencias F\'isico-Matem\'aticas,\\
Benem\'erita Universidad Aut\'onoma de Puebla, C.P. 72570,
Puebla, Pue., Mexico.}

\affiliation{Departamento de F\'{\i}sica, CINVESTAV, Apdo. Postal 14-740, 07000 M\'exico, D. F., M\'exico.}

\author{C. G. Honorato}
\email{carlos\_honorato@ymail.com}
\affiliation{Departamento de F\'{\i}sica, CINVESTAV, Apdo. Postal 14-740, 07000 M\'exico, D. F., M\'exico.}

\author{J. A. Orduz-Ducuara}
\email{jaorduz@gmail.com}
\affiliation{Facultad de Ciencias F\'isico -Matem\'aticas\\
Benem\'erita Universidad Aut\'onoma de Puebla, C.P. 72570, Puebla, Pue., Mexico.}

\author{M. A. P\'erez}
\email{mperez@fis.cinvestav.mx}
\affiliation{Departamento de F\'{\i}sica, CINVESTAV, Apdo. Postal 14-740, 07000 M\'exico, D. F., M\'exico.}

\pacs{14.80.Ec, 14.70.-e}


\date{\today}

\begin{abstract}
The general  two-Higgs doublet model (2HDM) contains a rich spectrum of neutral
and charged Higgs bosons,
whose detection would be a clear signal of new physics.
When the Higgs potential  is  CP conserving, the
spectrum includes a pseudoscalar mass eigenstate $A^{0}$,
which  does not couple to  vector bosons at tree level.
However, fermionic loops (top and bottom mainly) induce the coupling
$AVV'$ (with $V,V'=\gamma, Z$) at higher orders.
We evaluate the amplitude for the decays ${A^0} \to ZZ, Z\gamma, \gamma\gamma$, including
a generic fermionic loop contribution, and present results on
the branching ratios for 2HDM-I,II and III. Current LHC searches
on heavy Higgs bosons are used as an estimate to constrain the
allowed mass range for $A^0$.
\end{abstract}

\maketitle

\section{\label{sec:Intro} Introduction}

After many years of planning and preparation, the LHC has found evidence of a
Higgs-like  particle, with mass $m_h=125 \sim 126 \GeV$
\cite{higgs-atlas:2012gk,higgs-cms:2012gu}.
 It is remarkable that the observed Higgs mass falls within the range
preferred by the analysis of electroweak precision tests, within the
Standard Model \cite{Erler:2007sc}.
Although the measured  couplings point towards a SM Higgs
interpretation for such particle,
more data will be needed  in order to determine whether this resonance
belongs to the SM or to some of its extensions; in the later case
its properties could deviate from the SM expectations \cite{Gunion:1989we}.

On the other hand, the LHC has also searched for signals of new physics
beyond the SM, either through the production of new particles or by
looking for anomalous couplings for the SM particles \cite{Kane:2006hd}.
However, so far current LHC  studies have not detected any evidence of
new physics, and the resulting bounds on the associated scale has
been pushed into the TeV territory \cite{Arbey:2012bp}.
In fact, the weakest bounds are precisely on the search for
heavy Higgs particles \cite{Arganda:2013ve, Arganda:2012qp, Chakraborty:2013si},
which are predicted in many models of new physics,
including SUSY, XD, GUTs etc
\cite{Martin:1997ns, Quiros:2006my, Pomarol:2012sb, DiazCruz:2002er}.
Thus, searching for those Higgs particles could provide the
first signal on physics beyond the SM.  Furthermore, this
task could be attempted now with some degree of optimism,
because once the LHC has detected  a scalarlike state,
it seems possible that more
scalars could appear in the future LHC data.

One of the simplest extensions of the SM      consists of the addition of an
extra Higgs doublet,
the so-called two-Higgs doublet model (2HDM), which has been widely
studied in all the presentations that have been proposed
(2HDM I, II, III, X, Y, etc) \cite{Branco:2011iw}.
Some interesting properties of
the 2HDM include
\begin{itemize}
 \item A rich Higgs boson spectrum is predicted  within
 this model, which includes three neutral degrees of freedom and one
charged Higgs boson ($H^{\pm}$),
 \item Among the neutral states, the model predicts the existence of a
pseudoscalar state $A^0$,
 which would be a clear sign of new physics, and whose phenomenology
we are interested in.
\item When the Higgs potential is CP conserving,
$A^0$ is also a mass eigenstate \cite{Ginzburg:2004vp}
 \item Because of the quantum number assignments and discrete
symmetries of the model,
 this state does not couple to vector bosons at tree level. However,
such couplings could be induced at loop level
\cite{Mendez:1991gp, Bernreuther:2010uw}.
\footnote{
Besides presenting a numerical study
of  these loop-induced decays, Ref. \cite{Bernreuther:2010uw}  also presents an analysis  of
the reaction $pp \to VV$, but before data on $126 \GeV$ Higgs were
presented by the LHC.
}
\end{itemize}

In this paper, we are interested in studying the one-loop decays of the
pseudoscalar $A$ into a pair of vector bosons,
namely ${A^0} \to ZZ, Z\gamma, \gamma\gamma$,
within the context of the two-higgs
doublet model (2HDM), some of these decays have been studied in effective
Lagrangian context \cite{Perez:1995dc,DiazCruz:2001tn}. We shall work within the versions of the model
 where the Higgs sector  respects CP symmetry, which could occur in 2HDM-I, II and III;
in this case $A^0$ is actually a  mass eigenstate.
The loop amplitude for $A \to VV'$ receives contributions
from heavy fermions, mainly from the top and bottom quarks.

It turns out that the fermionic contribution within 2HDM I, II, depends
only on the Yukawa Lagrangian parameters,
which reduce in the end to $\tan\beta$ (the ratio of the
vacuum expectation values, i.e.
$\tan\beta=v_2/v_1$) and the fermion masses. On the other hand,
within the 2HDM-III, where one assumes some texture structure
for the Yukawa matrices \cite{Cheng:1987rs}, one needs to consider
additional parameters, which are called $\chi_{ij}$ \cite{DiazCruz:2004tr}.
For $i\neq j$ those couplings would induce
flavor-changing neutral currents (FCNC) mediated by the scalars,
 while for $i=j$ those couplings would
correct the usual 2HDM predictions for the diagonal Higgs-fermion
couplings \cite{Altmannshofer:2012ar}.
The dominant contribution to the loop amplitude in the low and
moderate $\tan\beta$ ($\simeq 1-5$)
comes from the top quark. For larger values of $\tan\beta$, which
seem disfavored by low energy
constraints on the 2HDM, the bottom quark contribution should also
be included.

The organization of this paper goes as follows.
Section \ref{sec:Model1} contains a
discussion of the general 2HDM and its
limiting cases, focusing on the Higgs-fermion couplings.
Section \ref{sec:proce} includes a discussion
of the decay amplitude for the process $A\to VV'$,
written in general terms, i.e. including the most general
couplings of the pseudoscalar ${A^0}$ with fermions; 
we also present the simplified
expressions for the decay widths of the
decays ${A^0} \to ZZ, Z\gamma, \gamma\gamma$.
Then, in  Sec. \ref{sec:resulta-analys} we discuss the numerical results
for the branching ratios, and we identify regions
of parameters where those decays show a large branching ratio.
Then, we study the constraints that current searches for heavy Higgs bosons
at the LHC could impose on the parameters of the model. This is done through
the evaluation of the signal strengths ($R_{ZZ}$), which are used as an estimate
for the signal coming from ${A^0} \to ZZ$. Our conclusions are
left  for Sec. \ref{sec:conclu}.

\section{\label{sec:Model1} The Two-Higgs doublet model (2HDM) }

In order to specify the 2HDM  versions, of types I , II and III, one
 needs to define the Yukawa sector,
which includes the interactions of the Higgs doublets with the quarks
and leptons. Interactions with
gauge bosons come from the covariant derivatives, and the pattern of
spontaneous symmetry breaking is associated with
the Higgs potential \cite{Ma:2006fn}. The general 2HDM-III
is defined by the Yukawa Lagrangian \cite{DiazCruz:2004pj}

\begin{eqnarray}
 {\dL}&=& {\yu}\overline{Q}_{L}^0 {\hgt} u_{R}^0  +
{{\yyuu}} \overline{Q}_{L}^0 {\hgtt} u_{R}^ 0
 + {{\yd}} \overline{Q}_{L}^0 {\hggs} d_{R}^0 +
{{\yydd}} \overline{Q}_{L}^0 {\hggss} d_{R}^0 + h.c.
\end{eqnarray}

\noindent
where

	\begin{eqnarray}
Q_L^0 = \left(
  \begin{matrix}
u_L \\[2mm]
d_L \\
  \end{matrix}
\right),
\hspace{3mm}
\overline{Q}_L^0 = \left( \overline{u}_L,\overline{d}_L \right),
{\hggs} = \left(
  \begin{matrix}
\phi_1^\pm \\[2mm]
\phi_1 \\
  \end{matrix}
\right),
\hspace{3mm}
{\hggss} = \left(
  \begin{matrix}
\phi_2^\pm \\[2mm]
\phi_2 \\
  \end{matrix}
\right),
\hspace{3mm}
\tilde{\Phi}_j &=& i \sigma_2\Phi_j^*=\left(
\begin{matrix}
{\phi_j}^* \\[2mm]-\phi_i^\mp \\
  \end{matrix}
\right),
\end{eqnarray}

\noindent
and $\phi_i = \frac{1}{\sqrt{2}} ( v_i + \phi^0_i + i\chi_i) $.

For the purposes of this paper, it suffices to consider the case
when the Higgs sector is CP
conserving, then the CP-even Higgs states ($h$ and $H$) come
from the mixing of the real parts of the neutral
components, $\phi^0_1 $ and $\phi^0_2$, while one combination of
the imaginary components, $\chi^0_1 $ and $\chi^0_2$, give place
to the pseudo-Goldstone boson (needed to give mass to the Z boson),
while the corresponding orthogonal combination denotes the
CP-odd state ${A^0}$.
 The mixing angles $\alpha $ and $\beta $ that appear in
the neutral Higgs mixing, corresponds to the standard notation,
i.e. $\tan\beta=v_2/v_1$.

Then,  the interactions of the pseudoscalar Higgs boson ($A^0$)
with the up-type quarks, are given
by the following Lagrangian \cite{DiazCruz:2010yq}:

\begin{equation}
\mathcal{L}_{up}^{neutral}=\overline{u}_{i}\left(S_{ijA}^{u}+%
i\gamma^{5}P_{ijA}^{u}\right)
u_{j} A^0 +h.c. \label{ygral}
\end{equation}

\noindent
with 

\begin{equation}
S_{ijA}^{u}=i\frac{\sqrt{m_{i}m_{j}}}{2\sqrt{2}v\cos \beta }\left(
\chi _{ij}-\chi _{ij}^{\dag }\right) ,  \label{svb3}
\end{equation}%
\begin{equation}
P_{ijA}^{u}=\frac{1}{2v}M_{ij}^{U}\cot \beta -\frac{\sqrt{m_{i}m_{j}}}{2%
\sqrt{2}v\sin \beta }\left( \chi _{ij}+\chi _{ij}^{\dag }\right)
\label{pvb3}
\end{equation}
Similar equations
hold for d-type quarks and leptons (see \cite{DiazCruz:2010yq}).

As discussed in Ref. ref \cite{DiazCruz:2004pj}, the assumption of
universal textures for the Yukawa matrices, allows to express
one Yukawa matrix, for instance $Y^f_2$, in terms of the quark masses, and parametrize
the flavor changing neutral scalar interactions (FCNSI) in terms
of the unknown coefficients
$\chi _{ij}$, which appear in the other Yukawa matrix, written in the mass-eigenstates
basis, namely
$\widetilde{Y}_{2ij}^{U}=\chi _{ij}\frac{\sqrt{%
m_{i}m_{j}}}{v}$, although other combinations are possible,
for instance the complementary textures discussed in Ref. \cite{Arroyo:2013kaa}.
These parameters can be constrained by considering all types
of low energy FCNC transitions; and although these constraints are
quite strong for transitions involving the first and second families,
as well as for the b-quark, it turns out that they are rather mild
for the top quark \cite{Buras:2010mh, Crivellin:2013wna}.

Furthermore, we only need to look at the diagonal couplings
of ${A^0}$ to up-, down-type quarks and charged leptons,
denoted generically as $f_i$, because of their contribution
to the loop amplitudes. Thus,
the relevant Lagrangian can be written as

\begin{equation}
\mathcal{L}^{f}_{A^0}=  \frac{gm^f_{i}}{2m_W}
\overline{f}_{i}\left(g^f_{Si}+%
i \gamma^{5}g^f_{Pi}\right)
f_{i} {A^0}.
\end{equation}

When the Yukawa matrices are taken to be Hermitian,
only the
pseudoscalar coupling remains; i.e., $g^f_{Si}=0$,
and one finds for the diagonal
coupling,

\begin{equation}
g^u_{Pi}=  \cot \beta -\frac{1}{\sin \beta } \left( \chi _{ii}\right),
\label{pvb3}
\end{equation}

\noindent
where the $\chi_{ii}$ can be taken essentially as  free parameters.

For 2HDM I and II, the $\chi$ vanish, and thus only the pseudoscalar
part contribute.
Table \ref{ta:vertex-I-II} shows the vertex ${A^0}f\bar{f}$ for $f=u,d$ type-quarks,
within the CP-conserving case.

\begin{table}[H]
\caption{\label{ta:vertex-I-II}
The vertex ${A^0}uu$ and ${A^0}dd$ for 2HDM-I-II and III
type \cite{Branco:2011iw}.
}
\begin{ruledtabular}
\begin{tabular}{c c c c}
           & Type I & Type II & Type III\\
  $g^u_P$  & $\cot\beta$ & $\cot\beta$& $\cot\beta-\frac{1}
{\sin\beta}(\chi^u_{ii})$\\
  $g^d_P$  &$-\cot\beta$ & $\tan\beta$& $\tan\beta-\frac{1}
{\cos\beta}(\chi^d_{ii})$\\
  $g^l_P$  &$-\cot\beta$ & $\tan\beta$& $\tan\beta-\frac{1}
{\cos\beta}(\chi^l_{ii})$\\
\end{tabular}
\end{ruledtabular}
\end{table}

\section{\label{sec:proce} The general expressions for the amplitudes and decay widths for $A \to VV'$}

In this section we shall present the calculation of the one-loop amplitude
for the decay $A\to VV'$, where $V,V'$ represent any neutral
SM vector boson ($V,V'=\{\gamma, Z\}$).
Due to the parity properties of the pseudoscalar,
the vertex $AV V'$ , as well as $AW^+W^-$,  are not present at tree level,
when the Higgs sector is CP conserving. However, this vertex could
be induced at one the loop-level from different
sources.  But in a model where the Higgs potential is CP conserving,
the coupling between $A^0$ and $H^+H^-$  is also forbidden,
and thus the charged Higgs does not contribute to the
loop-induced vertex $AVV'$. While in other models, such as the MSSM,
there are plenty of other particles that could contribute to the $AVV'$
vertex,  here we shall focus on the fermionic contributions only.
This choice is made because of our goal to perform a numerical
analysis based on  a few free parameters, as well as the recent
limits on the masses of new particles, beyond the SM, which are
reaching the TeV range, whose contributions to the vertex $AVV'$
are likely to be highly suppressed.
The  Feynman diagrams
for the amplitude
are shown in Fig. \ref{fd:A-ZZ-loop-level}. We shall consider the
most general ${A^0}f\bar{f}$ couplings, i.e. allowing for the
possibility of having a new source of
CP violating associated with the non-Hermicity of the Yukawa matrices.
\footnote{The numerical analysis for the case with CPV in the Higgs potential, and its
comparision with CPV from the Yukawa sector will be presented in a future publication.}
Then we shall present specific formulas for the decay widths within the
2HDM I, II and general III-type.

\subsection{The decay amplitudes for $A\to VV'$ }

Thus, the amplitude for the process $A \to VV',$  will be written in general,
namely we shall consider in equation (6).
The fermion-gauge vertices are written as:
$g_{Vff}^{} = -ik_{Vff}^{}\gamma_\mu(g_v^f - g_a^f \gamma^5),$
then for $V=Z$ we have $k_{Zff}^{} = \frac{g}{4\cos\theta_W}$
and for $V= \gamma$, $k_{\gamma ff}^{}=e |Q_f|\;, g^f_v=1\;, g^f_a=0$.

\begin{figure}[!htbp]
\centering
\includegraphics[scale = 0.7,angle = 0]{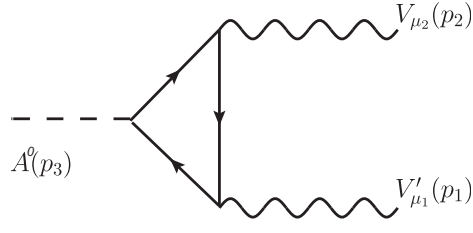}
\caption{Feynman diagram for $ A \to VV' $ decay, only fermion particles
are present. Crossed diagram is not shown.}
\label{fd:A-ZZ-loop-level}
\end{figure}

The kinematics conditions are defined according to the following
configuration of momentum:
$
p_3=p_1+p_2.
$
Then according to Fig. \ref{fd:A-ZZ-loop-level}, we have that:
$p_3^2=m_A^2, \;\;p_1^2=m_{V'}^2,
\;\;p_2^2=m_{V}^2$ and $2 p_1\cdot p_2=m_A^2-m_{V'}^2 -
m_{V}^2$.

The general tensorial amplitude for $AVV'$ vertex is written as follows,
\be
{\Mm}^{\mu_1\mu_2} = 
\frac{i  {g }\;r_f^{}  {N_C}\;k_{V_1ff} \; k_{V_2ff}}
{16 m_W  \pi ^2
\big(1-2  \big( {r_1}+
   {r_2}\big)+\big( {r_1}- {r_2}\big)^2\big)^2
}
{\Aa}_{VV'}^{\mu_1\mu_2} {\epsilon^*}_{\mu_1}
{\epsilon^*}_{\mu_2},
\ee

\noindent
where $r_i = \frac{M_{V_i}^2}{m_A^2},$
and

\be
{\Aa}_{VV'}^{\mu_1\mu_2} & = &
g_S^f
\Big(
{\Aa_1}
g^{\mu_1\mu_2}
+
{\Aa_2}\;
p_2^{\mu_1}  \;p_1^{\mu_2}
\Big)	
+
g_P^f\;
{\Aa_3}\;
\epsilon^{\alpha \beta\mu_1 \mu_2} p_{1_{\alpha}}
p_{2_{\beta}}.
\ee

\noindent
Here one can see how the  ${A^0}f\bar{f}$ couplings give place to
different tensorial structures, with the pseudoscalar part (i.e.
$g^f_{Pi}$)
inducing the term proportional to the Levi-Civita tensor, as
expected. The corresponding form factors are given by

\be
{\Aa_1} & = &
g_{v_1}^f g_{v_2}^f m_A^2\big(r_1^2-2 \big(r_2+1\big) r_1+
\big(r_2-1\big){}^2\big)\times
\nn\\&&\Bigg(\Bigg\{\frac{m_A^2}{2} \Big(4 r_f+2 r_1
\big(r_1 \big(4 r_f-r_2-3\big)-4 \big(r_2+2\big) r_f+r_1^2+r_2+3\big)-
1\Big)C_0({V_1},{V_2})
\nn\\&& + 2  r_1 (1-r_1+r_2) \Delta B_0(A,V_1)+2 r_1^2-2 \big(r_2+2\big)
r_1+1\Bigg\}+\Bigg\{1 \leftrightarrow 2\Bigg\}\Bigg)
\nn\\&& + g_{a_1}^f g_{a_2}^fm_A^2\big(r_1^2-2 \big(r_2+1\big) r_1+
\big(r_2-1\big){}^2\big) \times
\nn\\&&\Bigg(\Bigg\{\frac{m_A^2}{2} \Big(2 r_1^2 \big(-4 r_f-r_2-1\big)+
2 r_1 \big(8 r_f+r_2 \big(4r_f-1\big)-1\big)-4 r_f+2 r_1^3+1\Big)
C_0(V_1, V_2)
\nn\\&&+2 (2 r_2 r_1- r_1^2- r_2^2+2 r_1+2 r_2-1)B_0^{R}+ 2( r_2^2-
r_1 r_2-2 r_2- r_1+1)\Delta B_0(A, V_1)
\nn\\&&+2 r_1^2-2 \big(r_2+2\big) r_1+1\Bigg\}+\Bigg\{1 \leftrightarrow
 2\Bigg\}\Bigg),
\nn
\ee
\be
{\Aa_2} &= &   g_{v_1}^f g_{v_2}^f
\Bigg(\Bigg\{m_A^2  \big(r_1+r_2-1\big)C_0(V_1, V_2)\times
\nn\\&&\Big(2 r_1^2 \big(4 r_f-r_2-3\big)-2 r_1 \big(8 r_f+r_2  (4   r_f-5)-3\big)
+ 4 r_f+2 r_1^3-1\Big)
\nn\\&&+4\Big(r_1^3+4 r_2 r_1^2-2 r_1^2-5 r_2^2r_1+4 r_2 r_1- \big(r_1^2
-(r_2+2) r_1+2 r_2+1\big) r_1+ r_1\Big)B_0^R(A)
\nn\\&&+4(- r_1^3-4 r_2 r_1^2+2 r_1^2+5 r_2^2 r_1-4 r_2 r_1- r_1)\Delta
B_0(A, V_1)
\nn\\&&+4 r_1 \big(r_1^2-\big(r_2+3\big) r_1-r_2+3\big)-2\Bigg\}+
\Bigg\{1 \leftrightarrow 2\Bigg\}\Bigg)
\nn\\&&+\big(r_1+r_2-1\big) g_{a_1}^f g_{a_2}^f\times
\nn\\&&\Bigg(\Big\{m_A^2\Big(4 r_f+2 r_1 \big(r_1 \big(4 r_f-r_2-1\big)-
4 \big(r_2+2\big) r_f+r_1^2+3 r_2-1\big)+1\Big)C_0(V_1, V_2)
\nn\\&&+4\Big(r_1^2 -2 r_1 -r^2_2+2r_2\Big)B_0^R+4\big((r_2-1)^2+
(r_2+1)r_1-2r_1^2\big)\Delta B_0(A,V_1)
\nn\\&&+4 r_1^2-4 \big(r_2+2\big) r_1+2\Big\}+
\Big\{1 \leftrightarrow 2\Big\}\Bigg),
\nn
\ee

\noindent
and

\be
{\Aa_3} & = &
- m_A^2 g_{v_1}^f g_{v_2}^f\Bigg(\Big\{2 r_1^4-8 \big(r_2+1\big) r_1^3+
2 \big(3 r_2^2+4 r_2+6\big) r_1^2+4 \big(r_2-2\big) r_1+1\Big\}+
\Big\{1 \leftrightarrow 2\Big\}\Bigg)C_0(V_1, V_2)
\nn\\&&- g_{a_1}^f g_{a_2}^f\big(r_1^2-2 (r_2+1) r_1+(r_2-1)^2\big)
\times
\nn\\&&\Bigg(\Big\{m_A^2 \big(2r_1^2-2r_2 r_1-1\big)C_0(V_1, V_2)-4
\left(r_1-r_2+1\right) \Delta B_0(A, V_1)\Big\}+
\Big\{1 \leftrightarrow 2\Big\}\Bigg),
\nn
\ee

\noindent
where
$
B_0(i) = B_0(m_{i}^2,m_f^2,m_f^2),\;
\Delta B_{0} (i, j) = B_0(i) - B_0(j),\;
C_0(i, j)  =  C0(m_A^2, m_{i}^2,m_{j}^2,m_{f}^2,m_{f}^2,m_{f}^2).
$
We have used the renormalization method described in \cite{Pittau:2012zd},
which allows us to write:
$B_0^R = B_0(m_A^2,m_f^2+\mu_R^2,m_f^2+\mu_R^2)-
B_0(0,\mu_R^2, \mu_R^2),$
where $\mu_R$ denotes a renormalization scale.
These expressions show the Bose symmetry  explicitly.

\subsection{The decay widths for  ${A^0} \to ZZ, Z\gamma, \gamma \gamma$ }

In this section we shall present the expressions for the decay widths
corresponding to the processes: ${A^0} \to ZZ, Z\gamma, \gamma \gamma$,
which follow from the above expressions for the amplitudes.

\begin{enumerate}
 \item The expression for the decay width for ${A^0} \to ZZ$ is

\be
\Gamma({A^0} \to ZZ)
& = &m_A
\frac{ \kappa_{ZZ}^{} \,r_f^{}}{ (1 -4 r_Z^{})^{1/2}}\Bigg(	
\frac{{g_S^f}^2}{(1-4 r_Z^{})^3}
\Bigg(
{g_a^f}^4 {\Gg}_{a}^{ZZ}+ 2 {g_v^f}^2 {g_a^f}^2
{\Gg}_{av}^{ZZ}+{g_v^f}^4
{\Gg}_{v}^{ZZ}
\Bigg)
+
\nn\\&&
\frac{{g_P^f}^2}{2 }
\bigg(
{g_a^f}^2
\mathcal{F}_a^{ZZ}
-
{g_v^f}^2
\mathcal{F}_v^{ZZ}
\bigg)^2
\Bigg),
\ee

\noindent
where $\kappa_{ZZ}^{} = \frac{({N^f_C})^2}{64 \pi^5}
\Big(\frac{g m_f^{}}{2 M_W}\Big)^2
\Big(\frac{g}{4\cos\theta_W}\Big)^4.$
The ${\Gg}$'s and ${\mathcal{F}}$'s functions contain
Passarino-Veltman functions and also depend on
the ratios $r_{f}^{}$ and $ r_{Z}^{}.$

\item The decay width for ${A^0}\to Z\gamma$ is given by the following
expression:

\be
\Gamma({A^0} \to Z \gamma)
&=&
m_A
\frac{{g_v^f}^2  \kappa_{Z\gamma}^{}{r_f^{}}}{(1-r_Z^{})}
\bigg({g_P^f}^2 m_A^4 ({r_Z^{}}-1)^4{C_0(m_Z,0)}^2 +
\nn\\&&
2 {g_S^f}^2
\Big( {m_A}^2 (1-{r_Z^{}}) (4 {r_f^{}}+{r_Z^{}}-1){C_0(m_Z,0)}+
2 ({\Delta B_0(m_A, m_Z,0)}-1) {r_Z^{}}+2\Big)^2\Bigg),
\ee

\noindent
where $\kappa_{Z\gamma}^{}=\frac{({N^f_C})^2}{64\pi^5}
\Big(\frac{g m_f}{2 m_W}\Big)^2
\Big( \frac{g}{4 \cos\theta_W}\Big)^2
\Big(e |Q_f|\Big)^2$.

\item The decay width for ${A^0}\to \gamma\gamma$  is given by
\begin{eqnarray}
\Gamma({A^0} \to \gamma\gamma)
&=&
m_A^{} {r_f^{}}
\kappa_{\gamma\gamma}^{}
\Big(  I^2_f {g_P^f}^2 +2 {g_S^f}^2 \big(  I_f (4{r_f} - 1)+2 \big)^2 \Big),
\end{eqnarray}

\noindent
where $I_f= C_0(0,0) m_A^2$ and $\kappa_{\gamma\gamma}^{} = \frac{({N^f_C})^2 }
{128\pi^5}
\Big(\frac{g m_f^{}}{2 m_W^{}}\Big)^2
\Big(e |Q_f|\Big)^4$

\end{enumerate}

\section{\label{sec:resulta-analys} Results and LHC analysis}

The recent LHC results have shown that the observed Higgs
boson properties are very similar to the ones predicted by the SM,
although some small  deviations have persisted, which would
suggest the possible presence of new physics effects.
Within the 2HDM, those new effects depend on
the mixing angles and the scale $\mu_{12}^{}$, and thus in order to
get  small deviations with respect to SM, we shall choose the
following set of parameters,
\begin{eqnarray}
\mu_{12}^{}&=&200 \GeV\sim v,\\
\beta-\alpha&=&\frac{\pi}{2}+\delta,
\end{eqnarray}
where $v$ is the electroweak scale, and
$\delta$ is small. Thus, the above scenario remains
close to the SM limit.

\subsection{Numerical results for the branching ratios }

For the 2HDM of type II, the mass of the charged Higgs is constrained
to be above a value of order $350 \GeV$.
For the 2HDM of type-I  the charged Higgs mass
is less constrained, and it is possible to have a light charged Higgs,
and similarly for the 2HDM of type III \cite{Cordero-Cid:2013sxa}.
However, in order to explore a common scenario for 2HDM of type I,II and III,
we shall consider $m_{H^\pm}=350 \GeV.$

\begin{figure}[!htbp]
\centering
\includegraphics[width=2.2 in]{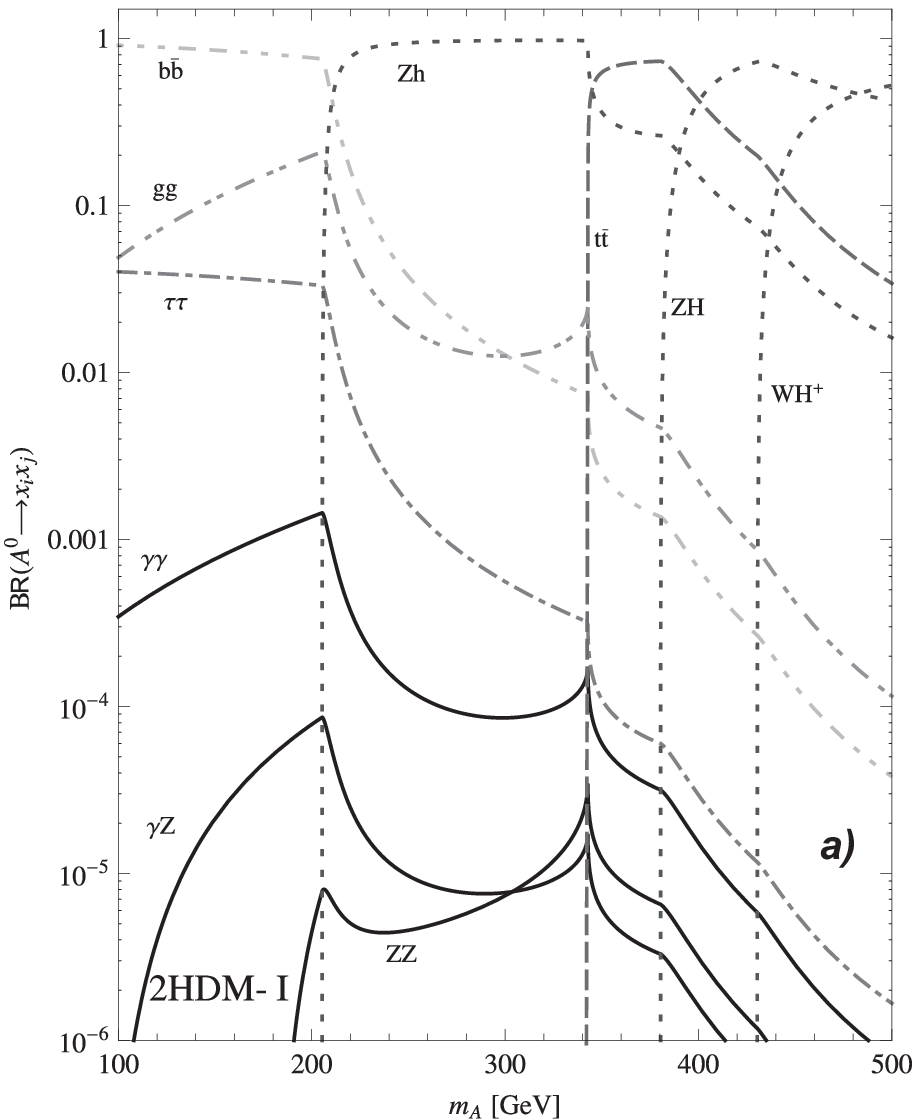}
\includegraphics[width=2.2 in]{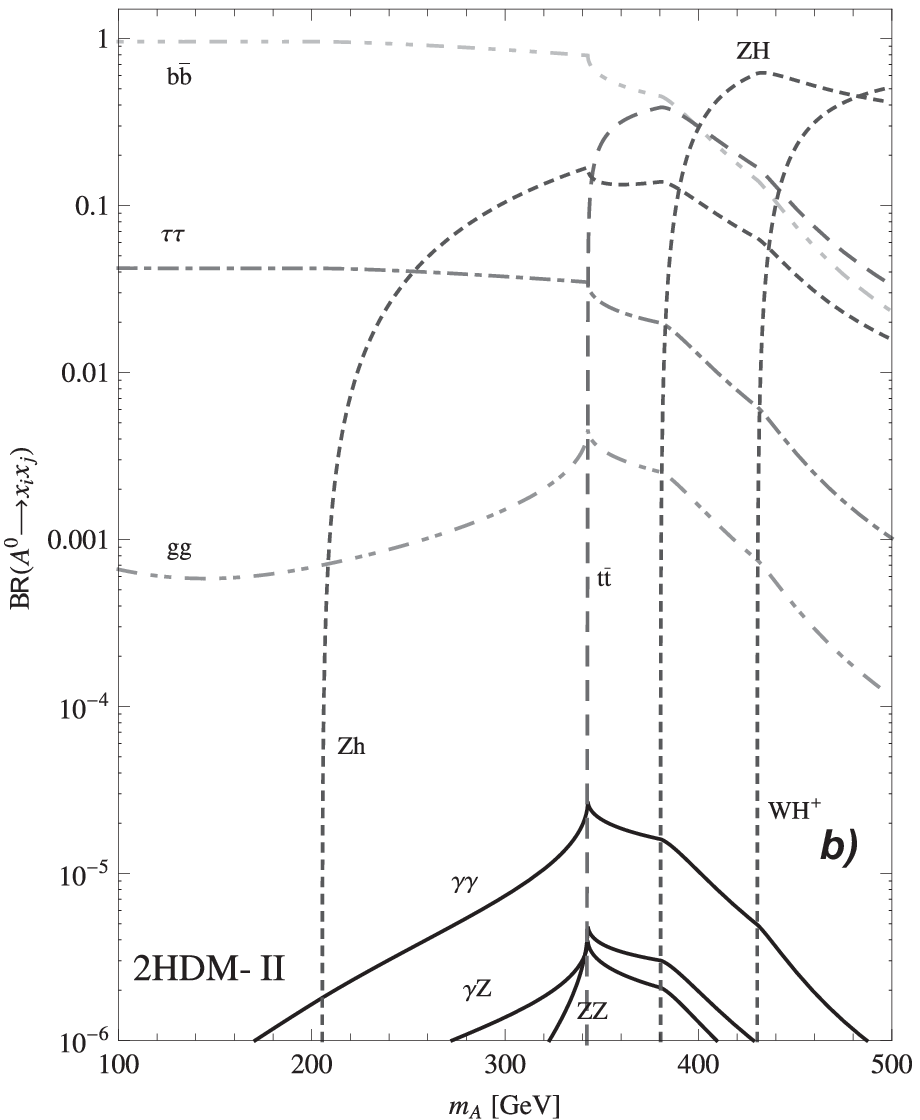}\\
\caption{Branching ratios for the pseudoscalar ${A^0}$
in 2HDM of type I and II. The parameter are:
$m_H=300 \GeV$, $m_h=125 \GeV$,
$m_H^\pm =350 \GeV,$
$\tan\beta=5$ and $\delta=0.1$.}
\label{All-Br}
\end{figure}

\noindent
In Fig. \ref{All-Br} we show the results for the branching
ratio of the pseudoscalar boson ${A^0}$ for the 2HDM of type I and II.
For THDM-I (see plot {\bf\small{a}} (left) in Fig. \ref{All-Br}), we can see that
whenever the channels $Zh, ZH, WH^\pm$
are kinematically allowed, they become dominant,
and the rest of the modes are suppressed, except for the decay into top quark
pair, which can dominate in a small window around 350-400 GeV.
The mode ${A^0} \to \gamma\gamma$ has a $\Br$ of order $10^{-3}$ in the best
case, for $m_A \simeq 200$ GeV, while the $\Br'$s for the modes $\gamma Z$ and $ZZ$ are
suppressed with respect to $\gamma\gamma$ by one and two orders of magnitude, respectively.
However, when the mass of $A$
is not enough to produce the final states $Zh, ZH, WH^\pm$,
the  modes $b\bar{b}, \; gg$ or even
$\tau\tau$  could become relevant.

For 2HDM-II (see plot {\bf \small{b}} (right) in Fig. \ref{All-Br})
the  modes $b\bar b$ and $ ZH $ are the dominant channels,
while the decay into gluons gets more
suppressed. In this case  the mode ${A^0} \to \gamma\gamma$
has a $\Br$ of order $2\times 10^{-5}$ , at most, for $m_A \simeq 350$ GeV,
while the BR for the  modes $\gamma Z$ and $ZZ$ is about
one order of magnitude smaller.

\begin{figure}[!htbp]
\centering
\includegraphics[width=2.4 in]{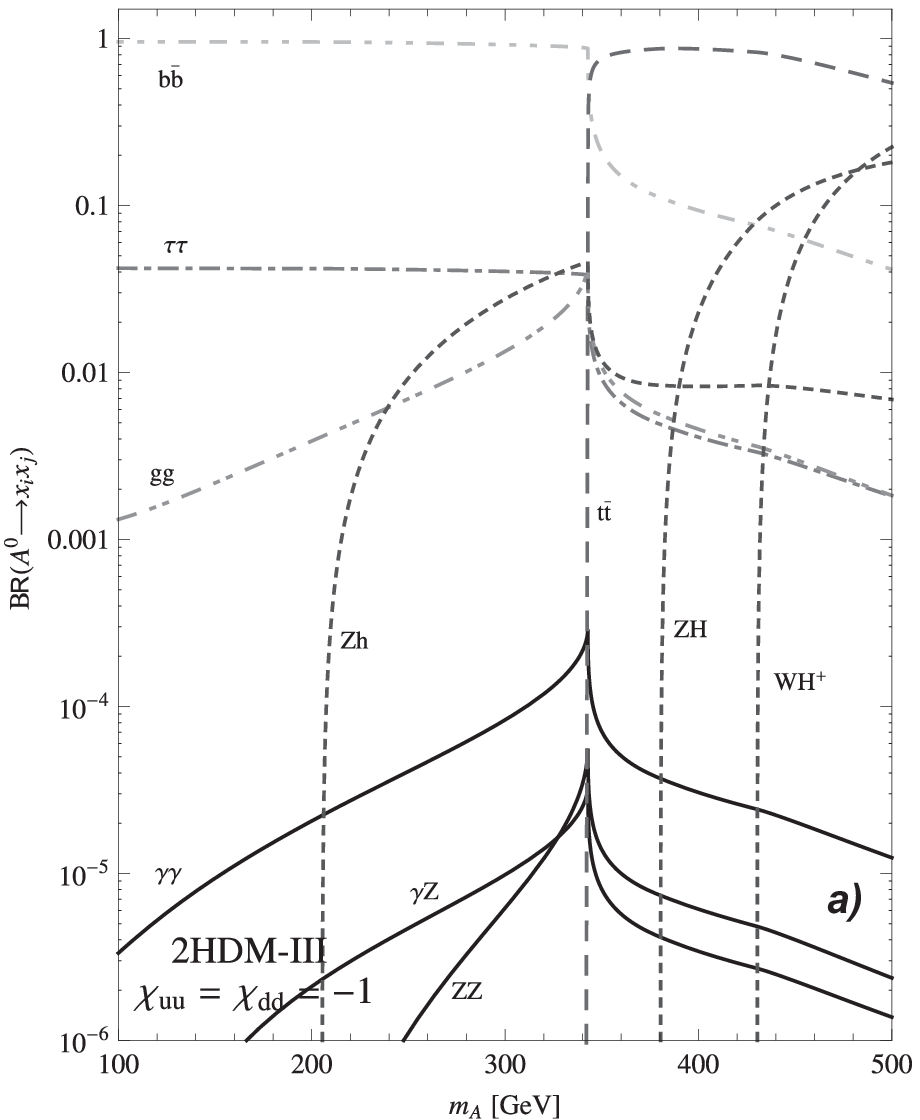}
\includegraphics[width=2.4 in]{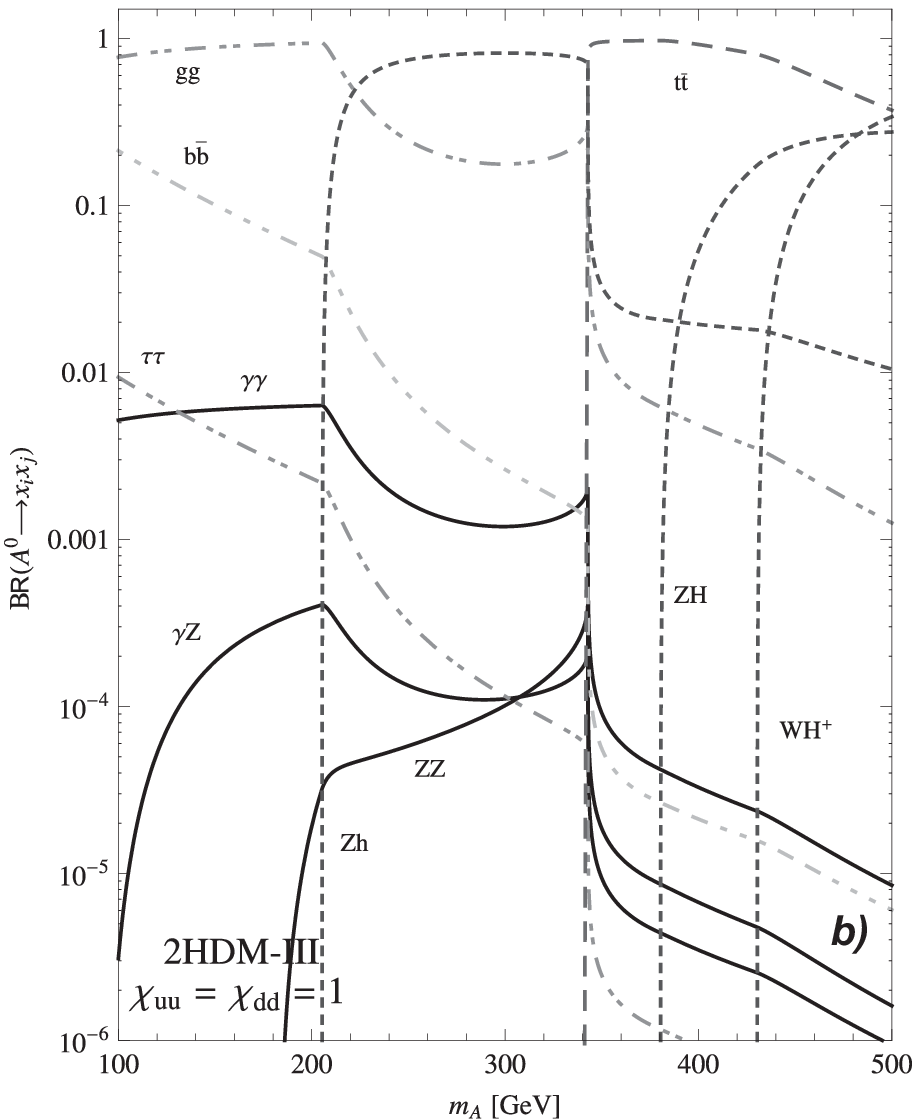}\\
\caption{Branching ratios for the pseudoscalar ${A^0}$ within 2HDM of type III in the
$CP$-conserving limit. The parameter are choosen as: $m_H=300 \GeV$, $m_h=125 \GeV$,
$m_{H^\pm}=350 \GeV$, $\tan\beta=5$ and $\delta=0.1$.}
\label{All-Br-III}
\end{figure}

In Fig. \ref{All-Br-III} we present the results for the $\Br$ corresponding to
the 2HDM of type III, in the $CP-$Conserving limit. But even in this  case the
Yukawa couplings are different with respect to the models with $Z_2-$symmetry,
as it was shown in table \ref{ta:vertex-I-II}.
In plot \textbf{\small a} (left)  we considered $\chi_{ff}=-1$, and for
a light boson ${A^0}$ the most importan channel is ${A^0}\to b\bar b$.
In this case we find $\Br({A^0}\to\gamma\gamma) \simeq 2\times 10^{-4}$
for $m_A \simeq 350$ GeV. For the same mass, the modes ${A^0}\to \gamma Z, ZZ$
have BR's of order $10^{-5}$.
 On the another hand, in  plot \textbf{\small b} (right),
we fix $\chi_{ff}^{}=1$, and this choice
significantly affects the channels ${A^0}\to b\bar b$ and ${A^0}\to \tau\tau$,
 reducing them even by about one order of magnitude. For this reason,
the $\Br({A^0}\to gg)$ becomes the dominant one for low masses. But now the mode
 $\gamma \gamma$ gets enhanced, and can reach $\Br$ of order $6 \times 10^{-3}$.
The modes $\gamma Z$ and $ZZ$ are also enhanced, but have $\Br$ at most of order
$3\times 10^{-4}$.

\begin{figure}[!htbp]
\centering
\includegraphics[width=2.8 in,height=2 in]{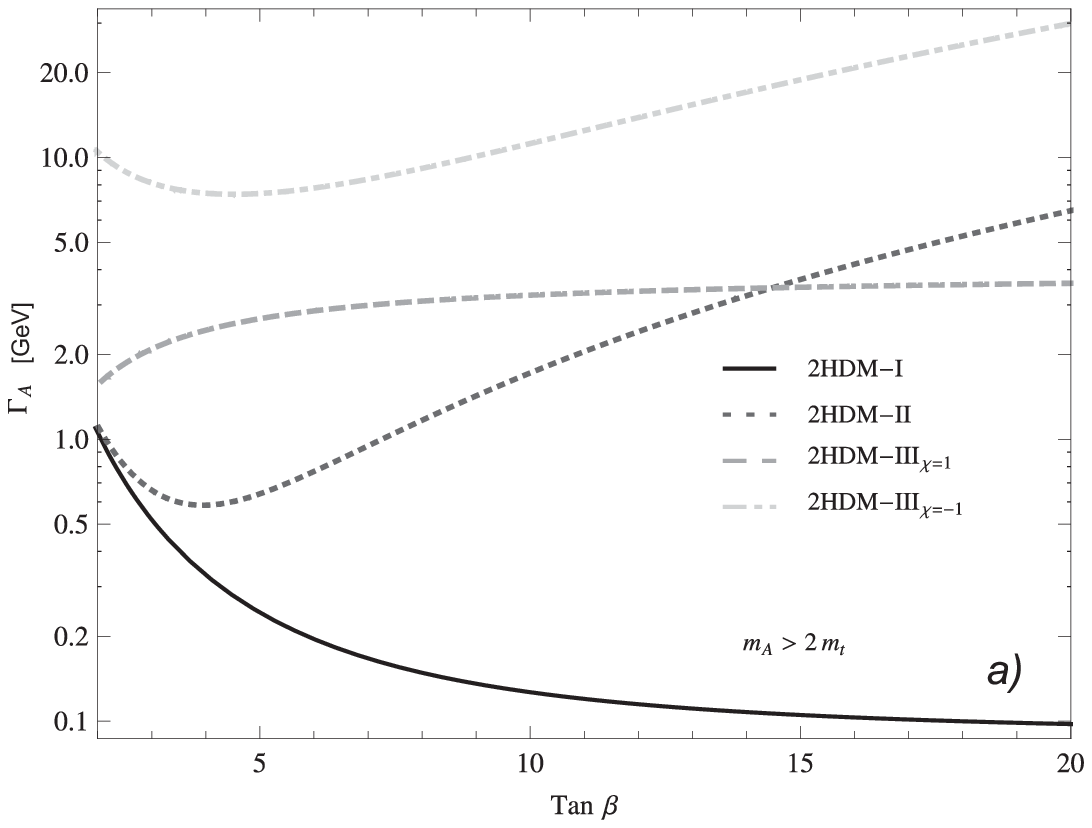}
\includegraphics[width=2.8 in,height=2 in]{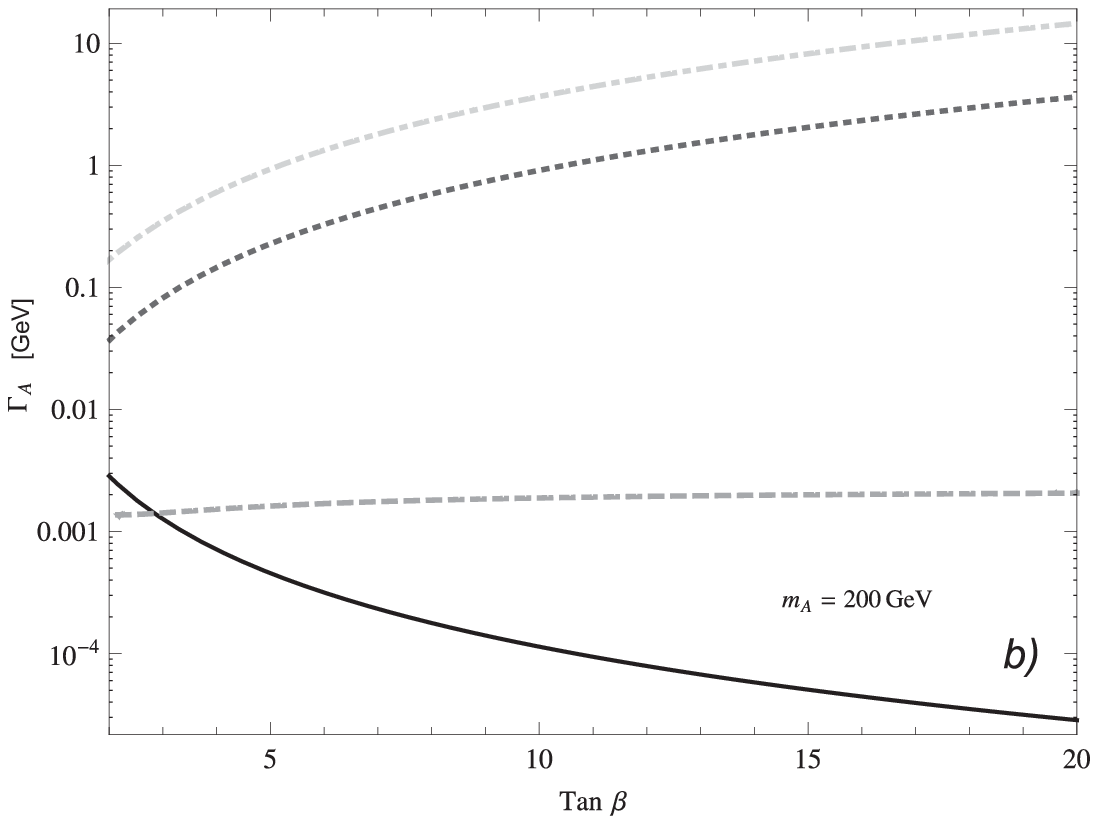}\\
\caption{Behavior of total width $\Gamma_A$ as a function of  $\tan\beta$.
The parameter are choosen as: $m_H=300 \GeV$, $m_h=125 \GeV$,
$m_{H^\pm}=350 \GeV$ and $\delta=0.1$.
The assignment of line codes appears in the plot {\bf\small{a}}, where we fixed
$m_A=355$ GeV ($>2m_t$), while in plot {\bf\small{b}} we take $m_A=200$ GeV. }
\label{G-Total}
\end{figure}

In order to understand the dependence of the
branching ratios on the parameters of the model, we present in
Fig. \ref{G-Total}, the total decay width for the pseudoscalar
$A^0$ state, as a function of  $\tan \beta$,  for the models
of type I, II and III.   In plot a) (left) we have chosen  a value
of $A$  mass slightly above the threshold for the
decay into top quark pair, namely $m_A= 355 > 2m_t,$,
while for plot b) (right) we have fixed
$m_A = 200 \GeV.$ We can see that for 2HDM-I, the total
width decreases with $\tan \beta$, because all fermionic
couplings go like $\cot \beta$, and this is so for both mass values of $A$.
For 2HDM-II  one notice that for $m_A=200$ GeV, the total width just
grows with  $\tan \beta$, a situation that reflects the fact that
the total width is dominated by the decay $A\to b\bar{b}$,
while for $m_A=200$ GeV the total width starts decreasing for low
$\tan \beta$ ($\simeq 1-4$),
but then increases with  $\tan \beta > 4$.
In this case, such behavior reflects the interplay between
the decays $A\to b\bar{b}$ and $A\to t\bar{t}$.
In the case of the 2HDM of type III, we observe a behavior of the total
decay width that
grows with $\tan \beta$, but with a more milder dependence.

\begin{figure}[!htbp]
\centering
\includegraphics[width=2.8 in,height=2 in]{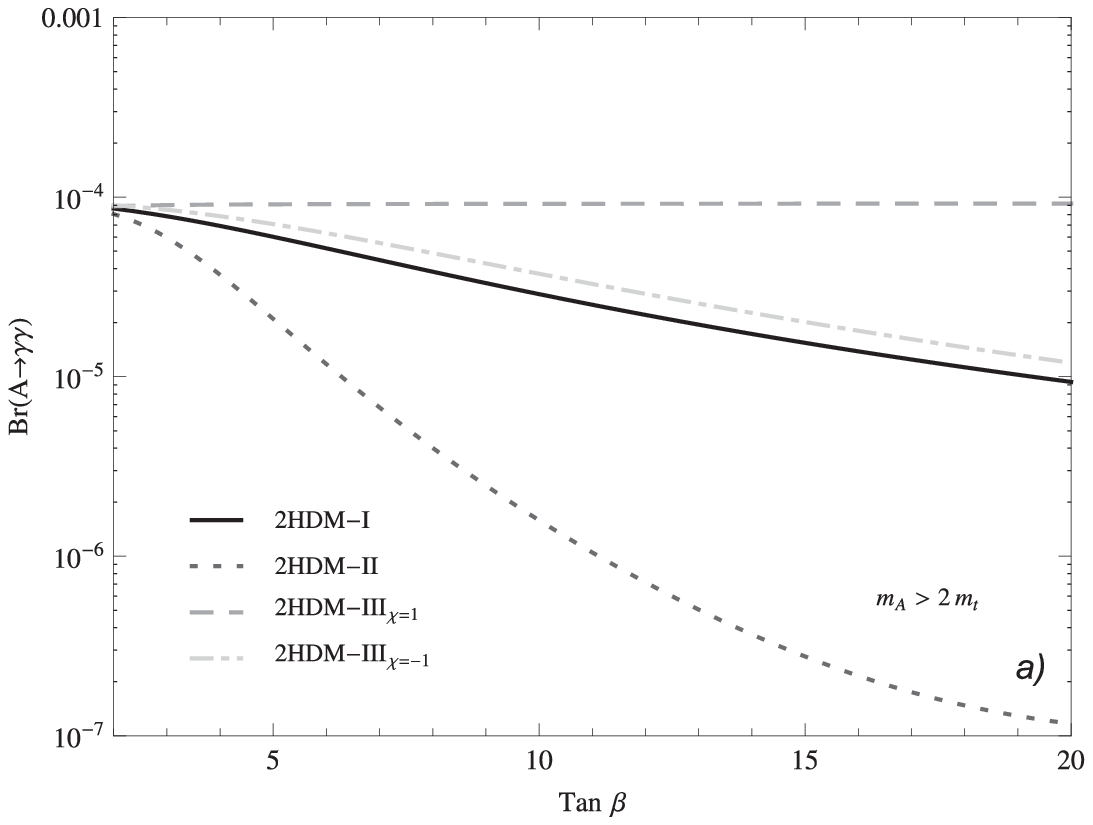}
\includegraphics[width=2.8 in,height=2 in]{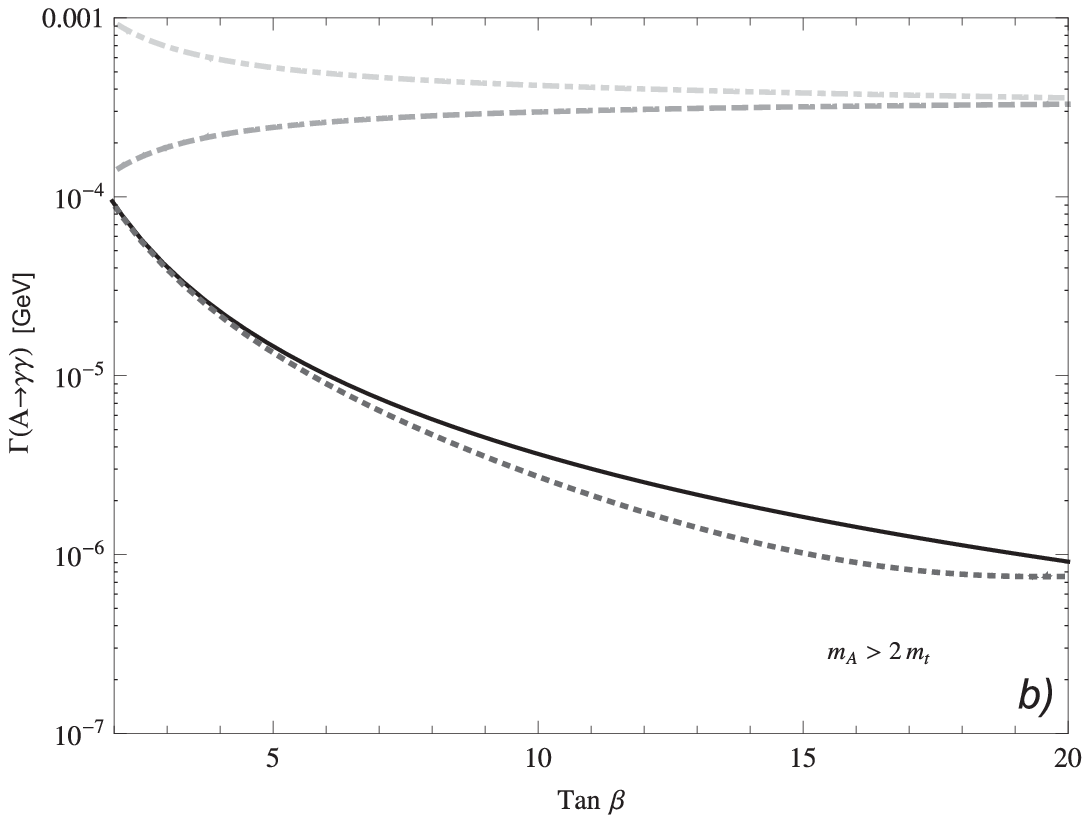}\\
\includegraphics[width=2.8 in,height=2 in]{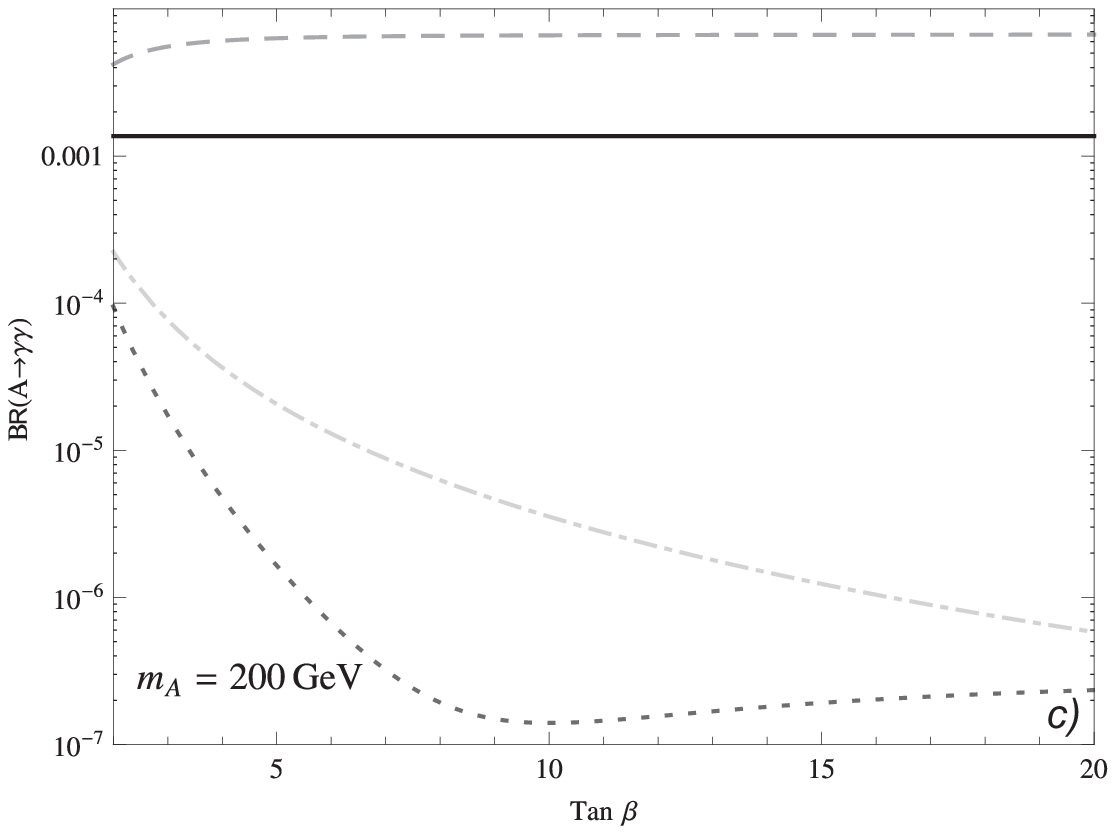}
\includegraphics[width=2.8 in,height=2 in]{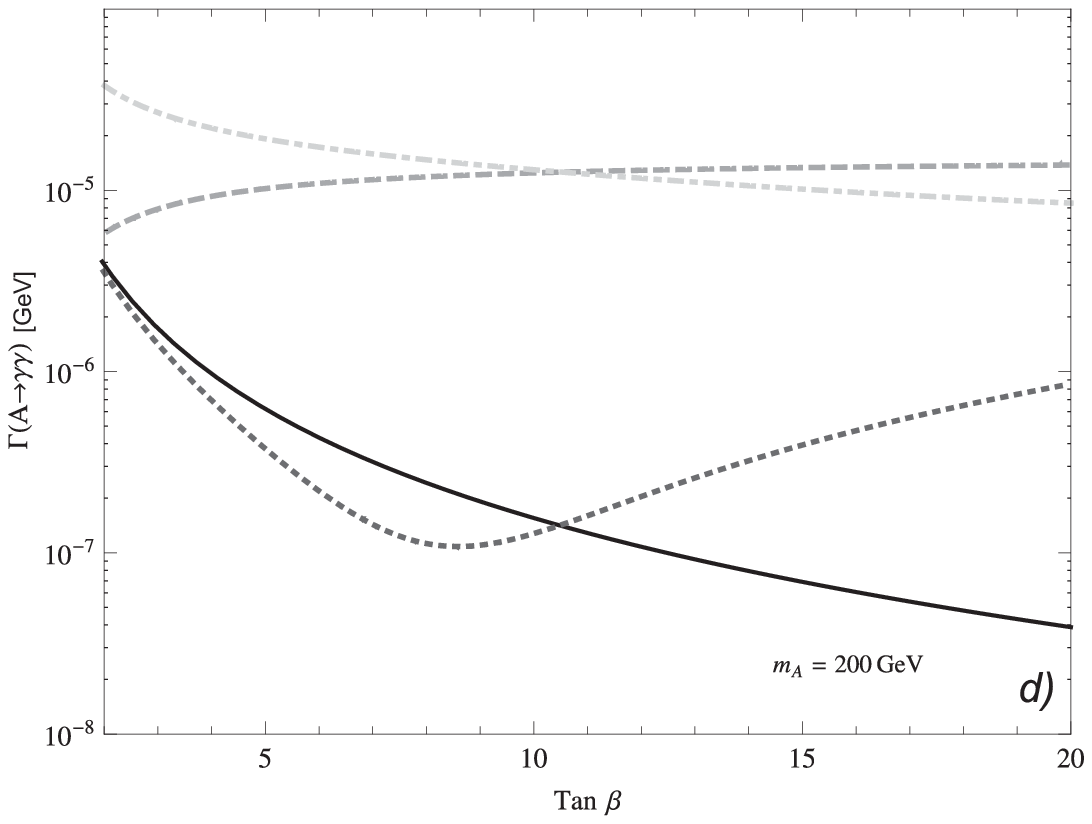}\\
\caption{Behavior of $\Br({A^0}\to \gamma\gamma)$ and $\Gamma({A^0}\to \gamma\gamma)$
as a function of  $\tan\beta$.
The parameter are chosen as $m_H=300 \GeV$, $m_h=125 \GeV$,
$m_{H^\pm}=350 \GeV$ and $\delta=0.1$.
The assignment of line codes appears in the plot {\bf\small{a}}. }
\label{Br-gg}
\end{figure}

In Ref. \ref{Br-gg}, we show the $\tan\beta$ dependence of
the BR into gamma pairs, as well as the corresponding partial width.
When one chooses a value $m_A=355$ GeV,
we can see that for 2HDM-I and 2HDM-II, the decrease of
$BR(A \to \gamma\gamma)$ with $\tan \beta$, just reflects
the corresponding behavior of $\Gamma(A \to \gamma\gamma)$;
while for 2HDM-III the dependence of $BR(A \to \gamma\gamma)$
on $\tan \beta$ comes from a combined effect of the
$\tan\beta$ dependence of the partial and total widths. On the
other hand, when we choose $m_A=200$ GeV,
within 2HDM-I $BR(A \to \gamma\gamma)$
remains constant, despite the fact that
 $\Gamma(A \to \gamma\gamma)$ decreases
with $\tan \beta$, but in this case the total width
 shows a similar suppression, which explains the
constant value of $BR(A \to \gamma\gamma)$.
Similar behavior is obtained for the modes $A \to Z\gamma$
and $A \to ZZ$. Overall,
one can see from these plots, that the 2HDM-II present the most
sensitive results,
showing a variation of about four orders of magnitude for
$\gamma\gamma$ and $\gamma Z$,
and more than four orders of magnitude for $ZZ$. In contrast,
the 2HDM-III with
$\chi_{ff}=1$ is less sensitive to $\tan\beta$; this scenario
presents small variations
(of order unity)  for all cases.

\begin{figure}[!htbp]
\centering
\includegraphics[width=2.8 in,height=2 in]{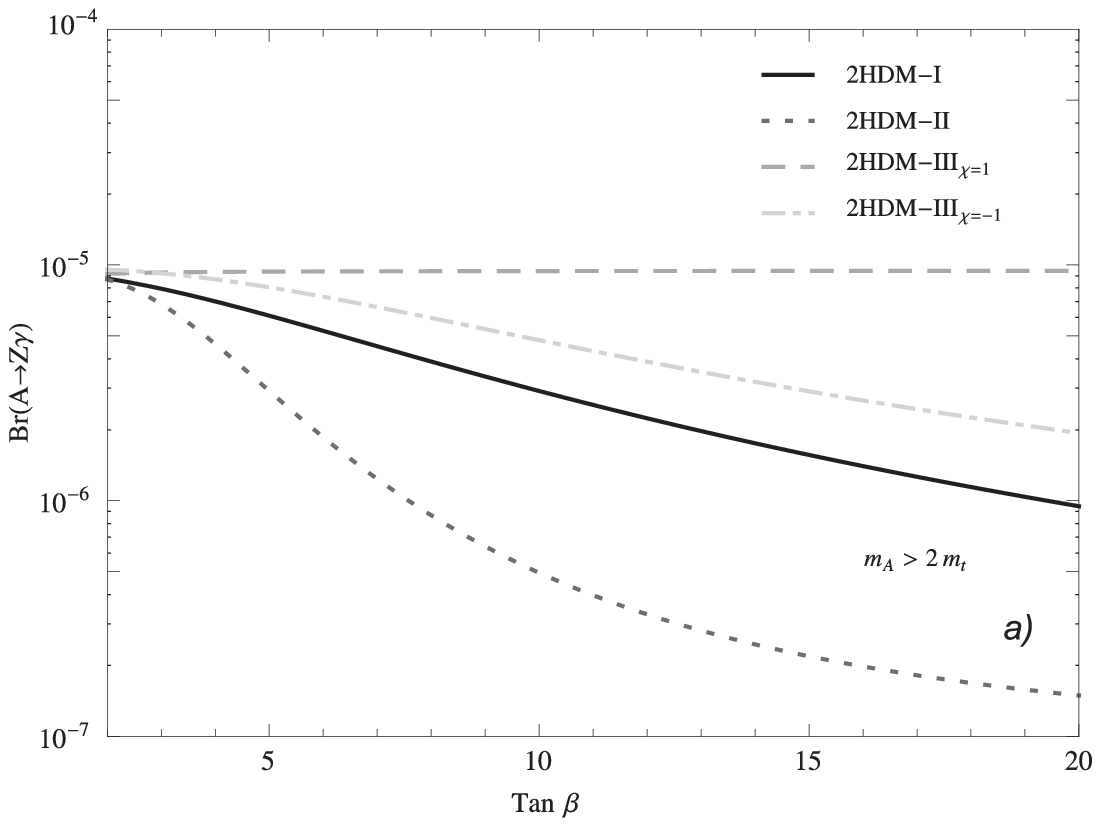}
\includegraphics[width=2.8 in,height=2 in]{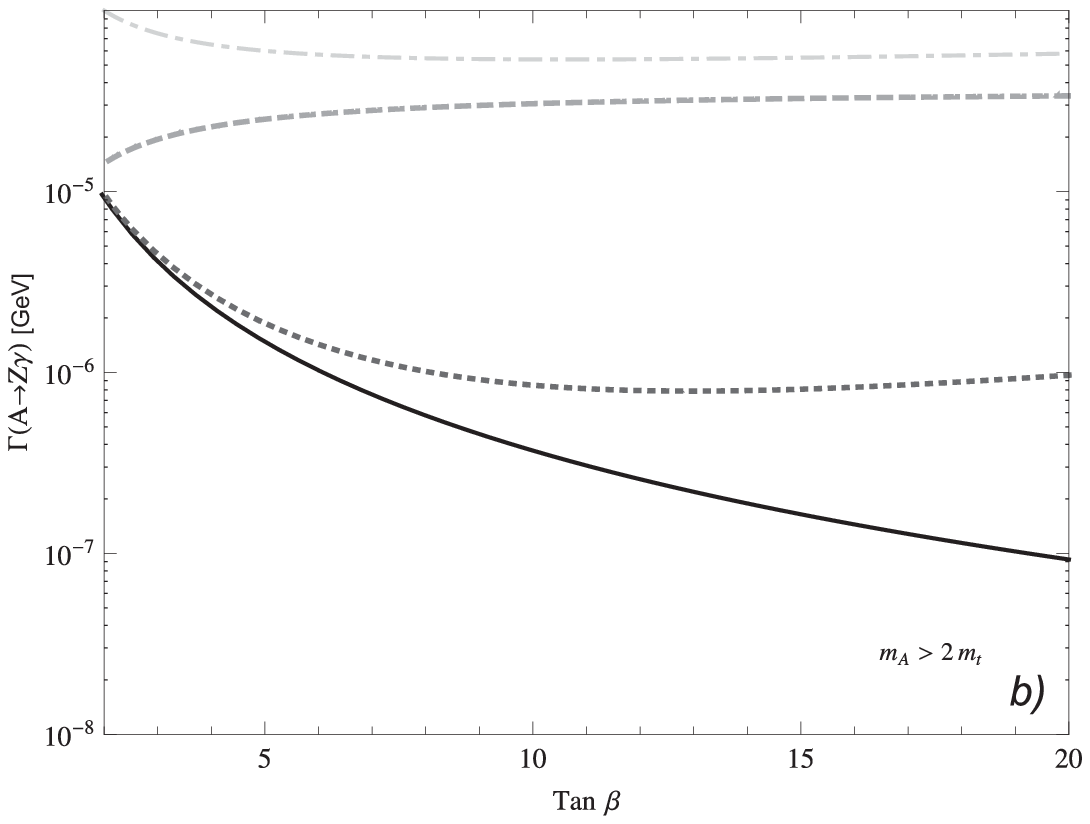}\\
\includegraphics[width=2.8 in,height=2 in]{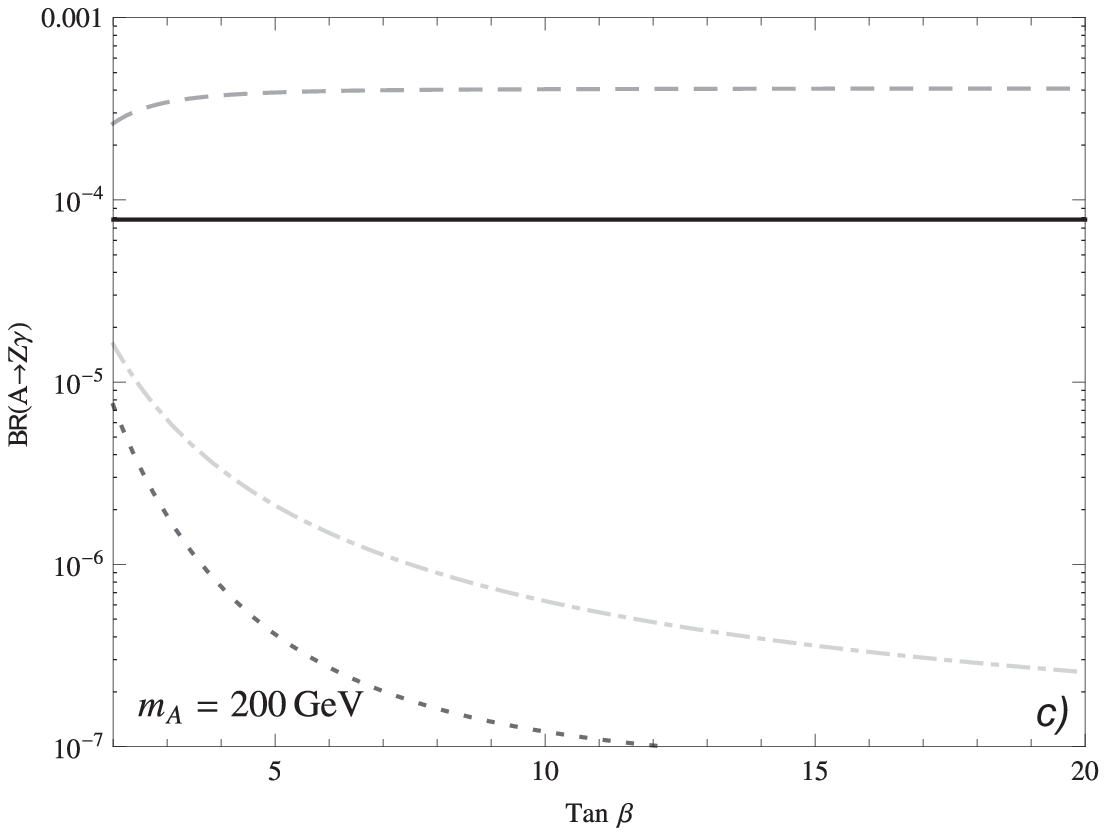}
\includegraphics[width=2.8 in,height=2 in]{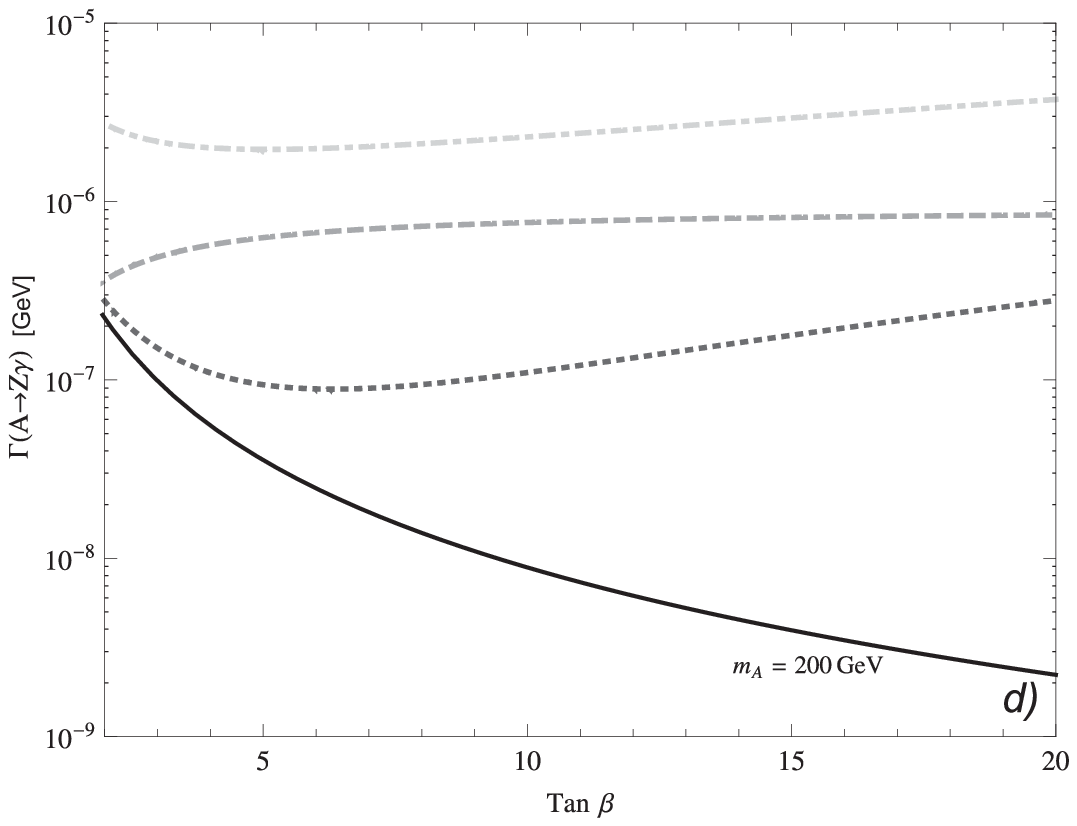}\\
\caption{Behavior of $\Br({A^0}\to Z \gamma)$ and $\Gamma({A^0}\to Z\gamma)$   as a function of  $\tan\beta$.
The parameter are chosen as: $m_H=300 \GeV$, $m_h=125 \GeV$,
$m_{H^\pm}=350 \GeV$ and $\delta=0.1$.
The assignment of line codes appears in  plot {\bf\small{a}}. }
\label{Br-gz}
\end{figure}

In Fig.\ref{Br-gz} we show the branching ratio for $A \to Z\gamma,$
and the corresponding partial decay width, as function of  $\tan \beta$.
For $m_A = 355 \GeV$, the ${\Br}$ for this mode   decrease with $\tan \beta$,
within 2HDM-I,  II and III (with $\chi =-1$), while within 2HDM-III with
$\chi =1$, the BR remains almost constant. On the other hand,  when one
chooses $m_A = 200 \GeV,$, we find that the $\Br$ decreases with
$\tan \beta$, within  2HDM-II and 2HDM-III (with $\chi =-1$), while
within  2HDM-I and 2HDM-III (with $\chi =1$), the $\Br$ remains almost
constant.

\begin{figure}[!htbp]
\centering
\includegraphics[width=2.8 in,height=2 in]{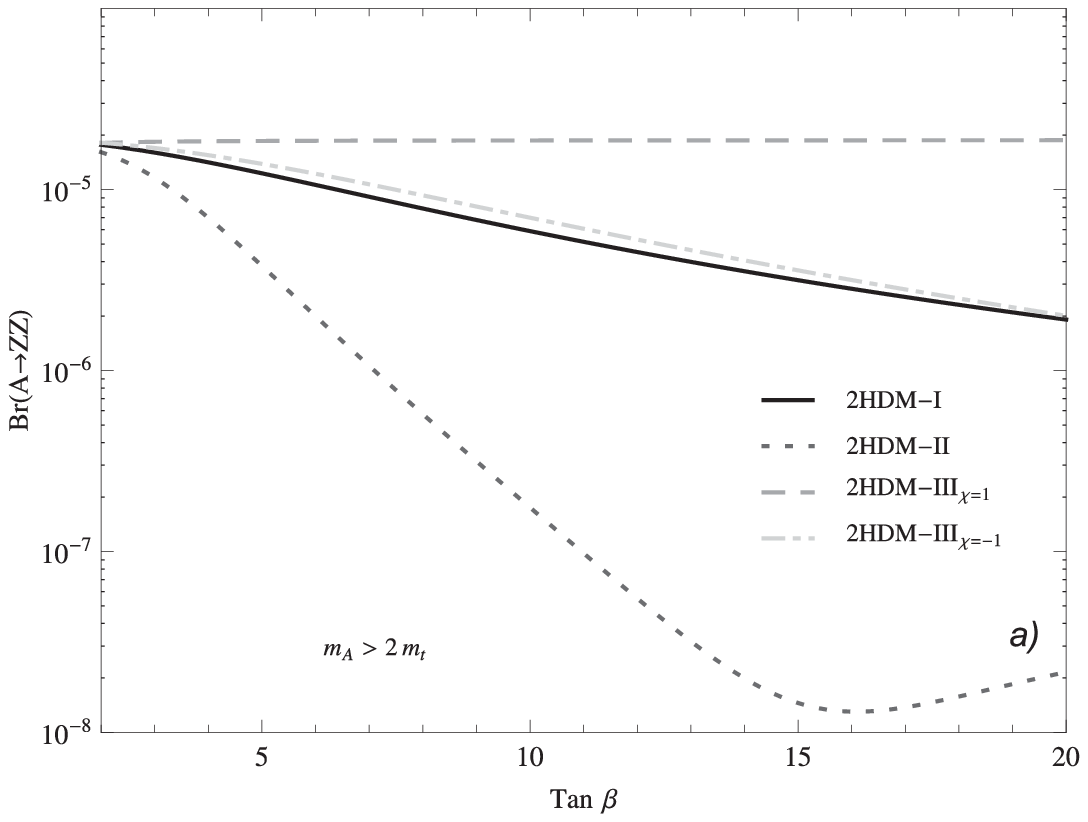}
\includegraphics[width=2.8 in,height=2 in]{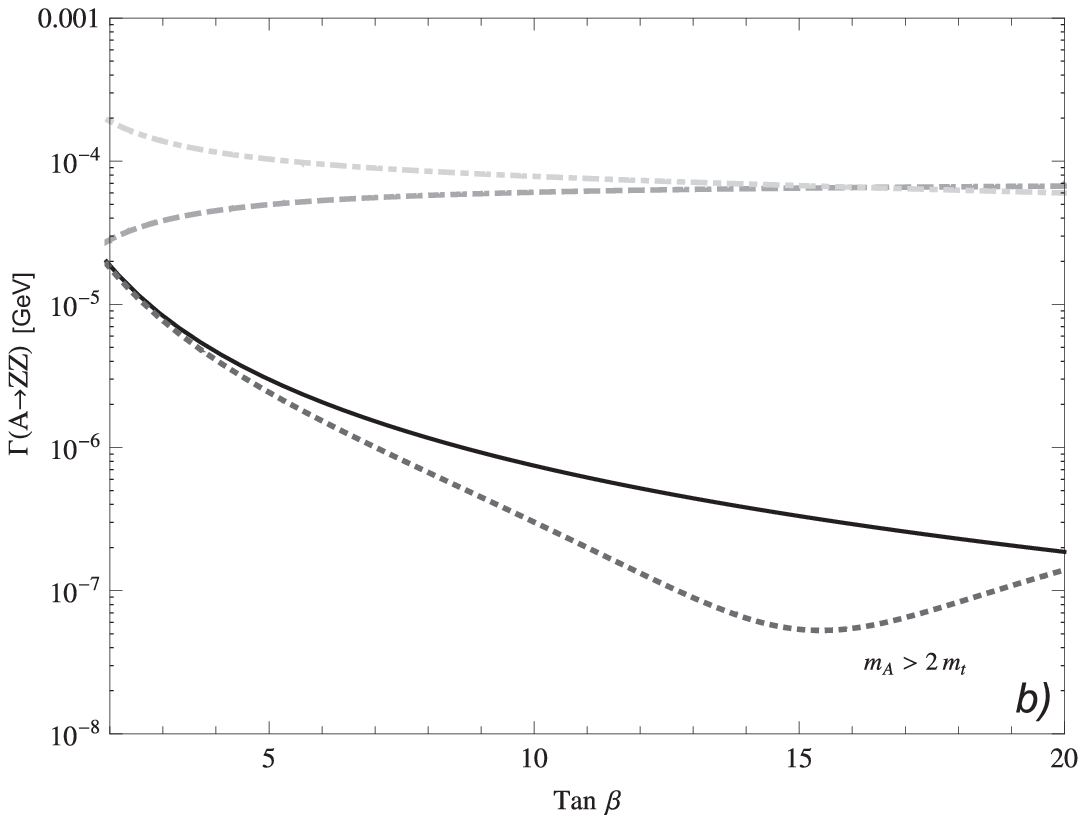}\\
\includegraphics[width=2.8 in,height=2 in]{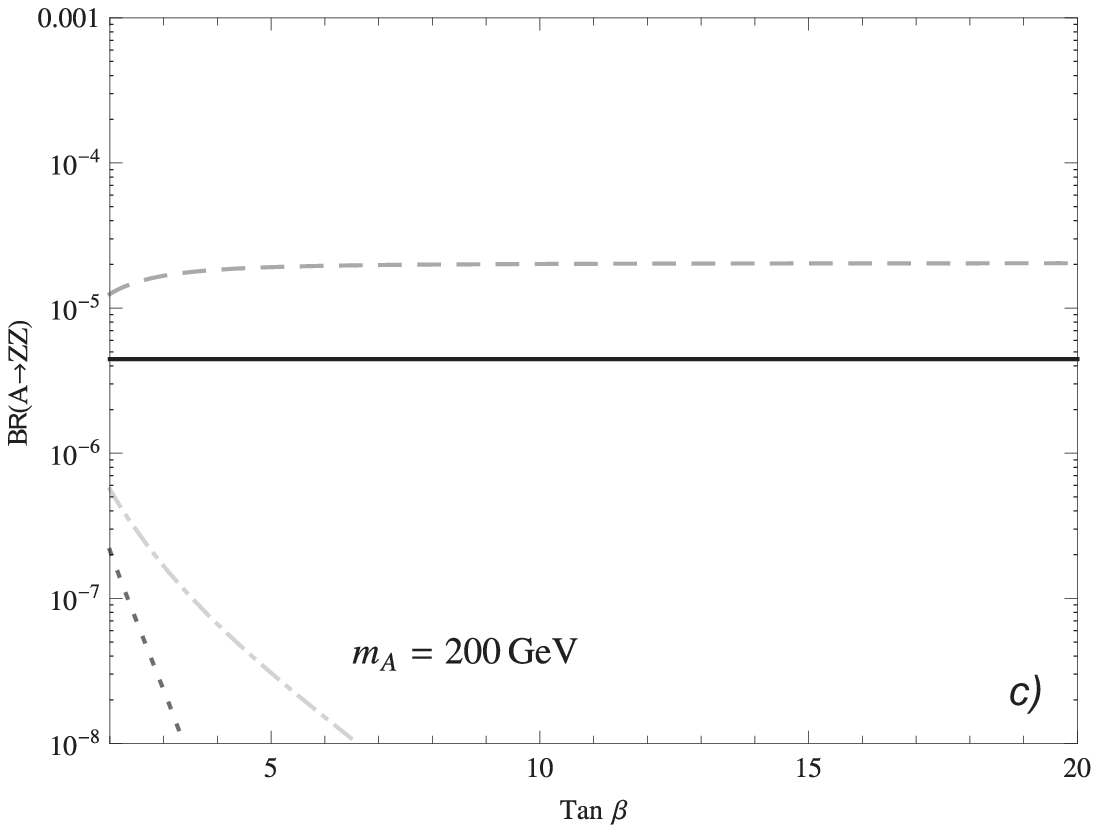}
\includegraphics[width=2.8 in,height=2 in]{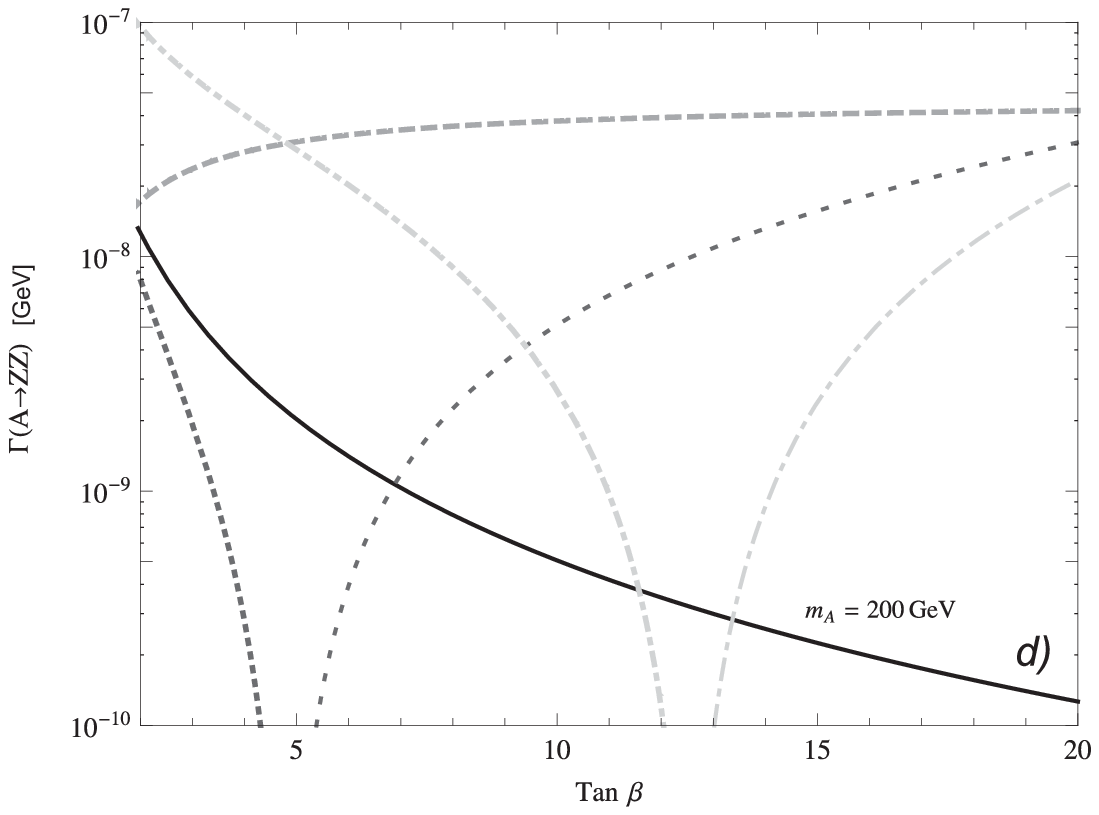}\\
\caption{Behavior of $\Br({A^0}\to Z Z)$ and $\Gamma({A^0}\to ZZ)$   as a function of  $\tan\beta$.
The parameter are chosen as: $m_H=300 \GeV$, $m_h=125 \GeV$,
$m_{H^\pm}=350 \GeV$ and $\delta=0.1$.
The assignment of line codes appears in  plot {\bf\small{a}}. }
\label{Br-zz}
\end{figure}

On the other hand, we show in Fig. \ref{Br-zz} the $\tan \beta$ dependence
of ${\Br}(A \to ZZ)$ and
$\Gamma (A \to ZZ)$. When $m_A = 355 \GeV$ the branching ratios decreases
as function of
$\tan \beta$ within  2HDM-I, II and 2HDM-III (with ${\chi = -1}$),
 while within 2HDM-III with $\chi =1$,
the BR remains almost constant.
When $m_A = 200 \GeV$ the BR decreasses with $\tan \beta$ within 2HDM-II and
III (with ${\chi =-1}$);
while within  2HDM-I and 2HDM-III (with $\chi =1$),
the $\Br$ remains almost constant. We also notice that
the $\Gamma(A \to ZZ)$ have a sharp decrease within 2HDM-II for
$\tan\beta \sim 5$ and
within   2HDM-III (with ${\chi = -1}$)  for $\tan\beta \sim 12,$ respectively.

\subsection{Constraints from LHC search for heavy Higgs bosons}

The first constraint that any extended model should fulfill nowadays
is the occurrence of a light
Higgs state with a mass near $m_{h^0}\simeq 125 \GeV$. After
considering the results, including
statistical and systematic uncertainties reported by
ATLAS and CMS \cite{higgs-atlas:2012gk, higgs-cms:2012gu}, we consider a central value for $m_{h^0}$
of 125 GeV and an uncertainty of $\pm 3 \GeV$; i.e., we accept a
value of $m_{h^0}$ in our numerical analysis if it lies within the range
[122-128 GeV]. Next, we also need to fullfil the  constraints coming from
the comparison with the SM-like Higgs signal observed at the LHC.

Thus, in order
to compare the signal rate observed for the SM-like Higgs signals, with
mass $m_{h^0} \simeq 125 \GeV$,
arising within the 2HDM  model, one can describe the signal
strength by the following ratios:

\begin{equation}
 R_{XX} = \frac{ \sigma( gg\to h^0 ) }{ \sigma( gg\to \phi_{sm} ) }
                \frac{ BR(h^0 \to XX) }{ BR (\phi_{sm} \to XX) }
\end{equation}
for $X=\gamma, Z$.

Within the so-called narrow-width approximation, we can
write the above expression for
$R_{XX}$ as follows:
\begin{equation}
 R_{XX} = \frac{ \Gamma(h^0 \to gg) }{ \Gamma(\phi_{sm} \to gg ) } \,
                \frac{ BR(h^0 \to XX)}{ BR(\phi_{sm} \to XX)}
\end{equation}

According to the CMS Collaboration the signal strength for the
$\gamma \gamma$ channel is
  $R_{\gamma\gamma} =0.78^{+0.28}_{-0.26}$,
while for the $ZZ$ channel it is $R_{ZZ}= 0.9^{+0.30}_{-0.24}. $
Thus, the light Higgs boson of the 2HDM, should satisfy the above
conditions, which is achieved in our scenarios because the properties of
the light Higgs boson were chosen to be very similar to the SM.

On the other hand, the LHC has also presented limits on the mass of a heavier
Higgs boson, which could be used in order to obtain some
constrains on the mass of the pseudoscalar state $A$.
We are aware that the pseudoscalar nature of $A$ will affect the
distributions of the particles appearing in the final states, and
strictly speaking those bounds that searched for the SM-Higgs can not be applied to
the pseudoscalar. However, we shall assume that those differences are small enough,
at least in order to obtain an estimate for the constraints on the  corresponding
mass.

For this purpose, we evaluate the ratio
\be
R_{XX} &=& \frac{\sigma (gg \to {A^0}) {\Br}({A^0} \to XX) }{\sigma
(gg \to \phi_{sm} ) {\Br}(\phi_{sm} \to XX)} \\
  &=& \frac{\Gamma ({A^0} \to gg) {\Br}({A^0} \to XX) }{\Gamma
( \phi_{sm} \to gg  ) {\Br}(\phi_{sm} \to XX)}
\ee
at a mass value $m_{\phi_{sm}} = m_A$, which we vary
over the range $180 < m_A < 360 \GeV$, in order to
stay below the threshold for the decay into top pair, which
becomes dominant then. The results are shown in the following Fig.
  \ref{fig:RXX-2HDM-III-RZZ}, which
shows the values of $R_{ZZ}$ and $R_{\gamma\gamma}$ vs $m_A$
for 2HDM-I, II and III. We have included the CMS exclusion contour for each channel
(ZZ and $\gamma\gamma$), and whenever the predictions from the models fall above these
lines, such scenarios would be excluded.

From the left figure, we can see that all models satisfy the
constraints imposed by the heavy Higgs
search in the ZZ channel. On the other hand, we can see from the right figure,
that the values of $R_{\gamma\gamma}$ bounded at LHC, could exclude the mass range
$100< m_A< 160$ GeV for the 2HDM of type III for the choice $\chi=1$.
The 2HDM of type II satisfy this constraint for the
mass range $180 < m_A < 360 \GeV$,
while 2HDM-I and 2HDM-III (with $\chi=-1$) seem to be excluded only
in the mass range $138 < m_A< 144$ GeV. These are promising results which deserve to be looked
at in more detail by the experimental LHC Collaborations.

\begin{figure}[!htbp]
\centering
\includegraphics[width=3 in]{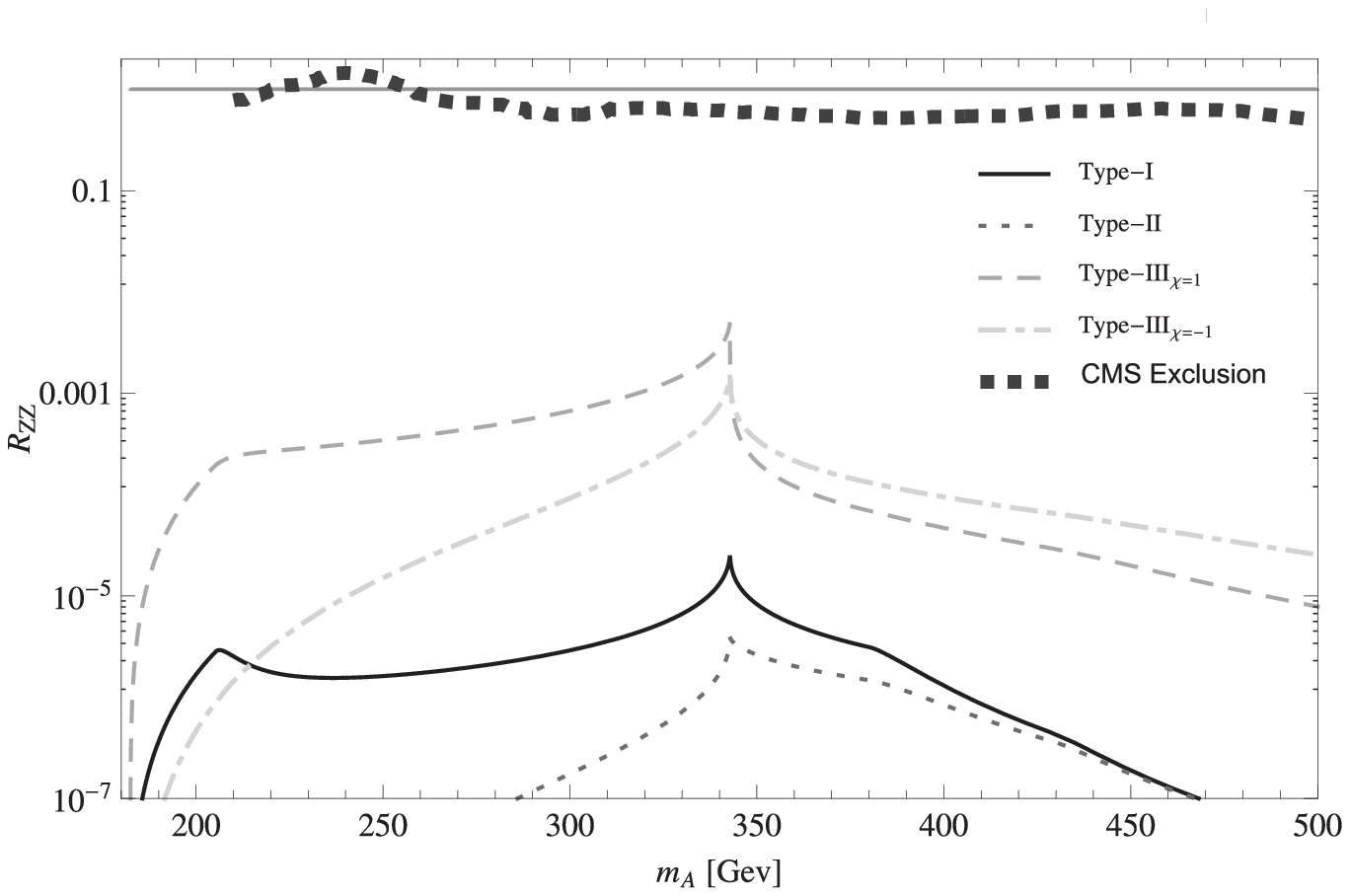}
\includegraphics[width=3 in]{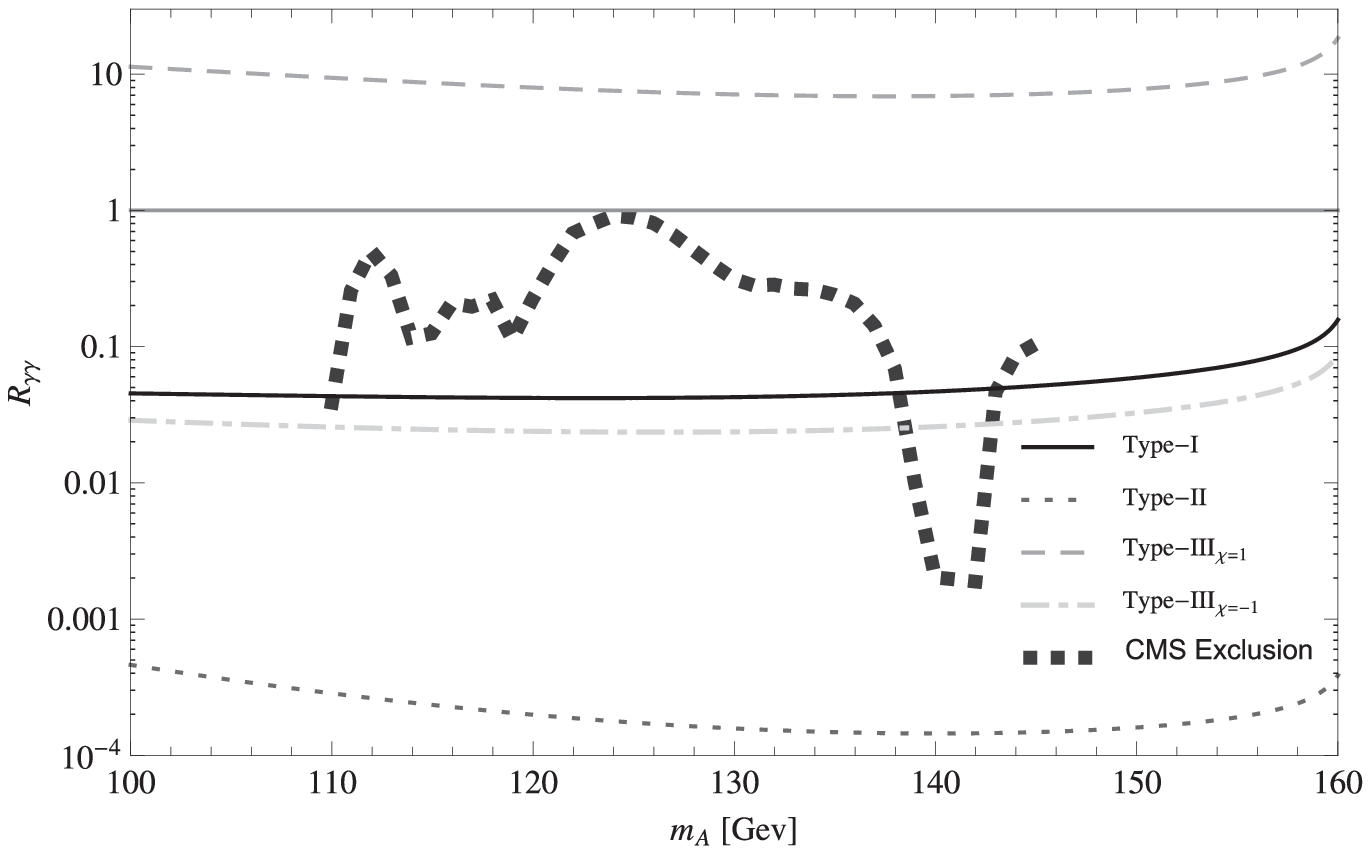}
\caption{
$R_{ZZ}$ and $R_{\gamma\gamma}$ vs $m_A$ in
2HDM-I-II and III.
\label{fig:RXX-2HDM-III-RZZ}}
\end{figure}

\section{\label{sec:conclu}
Conclusions}

As it is well known, the general  two-Higgs doublet model (2HDM) contains a rich spectrum
of neutral and charged Higgs bosons, whose detection at current and future
colliders would be a clear signal of new physics.
When the Higgs potential  is  CP conserving, the neutral
spectrum includes a pseudoscalar mass eigenstate $A^{0}$.
Even in this case, the interactions of ${A^0}$ with fermions could include
a CP-violating contribution, arising from a possible non-Hermiticity of the
Yukawa matrices. When the Higgs sector is CP conserving,
the ${A^0}$ boson  does not couple to  vector bosons at tree level.
However the coupling ($AVV'$) is generated at loop level, from
fermionic and bosonic loops.
The dominant contribution in the low and moderate
$\tan\beta$ ($\simeq 1-5$),
comes from the top quark, while for larger values of $\tan\beta$,
the bottom quark contribution becomes relevant.

We have evaluated the generic fermionic contribution to the decays
${A^0} \to ZZ, Z\gamma, \gamma\gamma$, including its scalar and pseudoscalar
vertices. Then, we have presented numerical results for
the branching ratios. We found that there are regions of parameters where
such loop-induced modes could reach significant branching ratios.
Current LHC searches for heavy Higgs bosons are used
to derive an estimated constraint on the parameters of the models.
We find that for 2HDM-II the whole mass range is acceptable, for our choices of parameters,
while the 2HDM-III with $\chi=1$ is excluded in the  mass range ($100 < m_A < 160$ GeV).
On the other hand,   2HDM-I and 2HDM-III (with $\chi=-1$) seem to be excluded only
in the mass range $138 < m_A< 144$ GeV.
These scenarios should be further studied at the LHC13 in order to
confirm the estimates for exclusion limits presented in this paper.

\begin{acknowledgments}
We acknowledge support from CONACYT-SNI (Mexico), and discussions with J. Montano, who
participated in the early stages of this work.
We would like to thank M. Wiebusch, \textit{et  al}. for bringing
Ref. \cite{Bernreuther:2010uw} to our attention after our paper was submitted
to E-prints.
\end{acknowledgments}

\end{document}